\documentclass{aastex}

\usepackage{psfig}
\usepackage{graphicx}
\usepackage{emulateapj5}
\usepackage{times}



\newcommand{\lsim}{\mbox{\hspace{.2em}\raisebox{.5ex}{$<$}\hspace{-.8em}\raisebox{-.5ex}{$\sim$}\hspace{.2em}}}
\newcommand{\gsim}{\mbox{\hspace{.2em}\raisebox{.5ex}{$>$}\hspace{-.8em}\raisebox{-.5ex}{$\sim$}\hspace{.2em}}}

\def\asca       {{\em ASCA}\/}
\def\chandra    {{\em Chandra}\/}

\def\rosat      {{\em ROSAT}\/}

\def\hst        {{\em HST}\/}
\def\mydegree{$^\circ\mskip-5mu$}

\usepackage{mathptmx}

\begin{document}


\title{A small X-ray corona of the narrow-angle tail radio galaxy NGC 1265 soaring through the Perseus cluster}

\author{M.\ Sun, D.\ Jerius, \& C.\ Jones}
\affil{Harvard-Smithsonian Center for Astrophysics, 60 Garden St.,
Cambridge, MA 02138;\\ msun@cfa.harvard.edu}

\shorttitle{An X-ray galaxy corona in NGC~1265}
\shortauthors{SUN ET AL.}

\begin{abstract}

A deep \chandra\ observation of NGC~1265 (3C 83.1B), the prototype
for the narrow-angled-tailed (NAT) radio galaxy, reveals a small
cool X-ray thermal corona ($\sim$ 0.6 keV) embedded in the hot ICM
of the Perseus cluster ($\sim$ 6.7 keV). The corona is asymmetric
with a sharp edge ($\sim$2.2$''$, or 0.8 kpc from the nucleus) to the
south and an extension to the north (at least $\sim 8''$ from the
nucleus), which is interpreted as the action of ram pressure while
solely the static ICM confinement is unable to explain.
We estimate that the corona is moving with a velocity of $\sim$
2.4 - 4.2 times the local sound speed to the south. The
presence of the sharp edge for this small corona indicates that the
transport processes are largely suppressed by the magnetic field
there. The magnetic field around the corona also suppresses heat
conduction by at least a factor of $\sim$ 60 across the corona boundary.
We conclude that it is unrealistic to study the interaction of the
small X-ray coronae with the hot ICM without the consideration of
the roles that magnetic field plays, a factor not included in
current simulations. An absorbed (N$_{\rm H}$ = 1.5 - 3 $\times$
10$^{22}$ cm$^{-2}$) nucleus is also detected, which is not usual
for FR I radio galaxies. Weak X-ray emission from three inner radio
knots in the jets is also detected. Indentations
at the east and west of the corona indicate interaction between the
jets and the X-ray corona. Narrow jets carry great amounts of energy
out of the central AGN and release the energy outside the corona,
preserving the tiny and vulnerable corona. This case reveals that
the inner kpc core of the corona of massive galaxies can
survive both high-speed stripping and powerful AGN feedback. Thus,
the cooling of the X-ray coronae potentially provides fuel to the
central SMBH in rich environments where the amount of the galactic
cold gas is at a minimum.

\end{abstract}


\section{Introduction}

\chandra\ observations have discovered eight small but spatially resolved
thermal coronae of early-type galaxies in hot cluster (kT$_{\rm ICM} >$
3 keV) environments (Vikhlinin
et al. 2001; Yamasaki, Ohashi \& Furusho 2002; Sun et al. 2005, S05 hereafter).
As summarized in S05, the coronae of massive galaxies can survive in spite
of the evaporation and gas stripping by the hot and dense ICM, as well as the
energy imparted by the central AGN outbursts. Thus, the properties of these
``survivors'' can put constraints on important physics, e.g., microscopic
transport processes, gas stripping, stellar mass injection and AGN heating
(e.g. Vikhlinin et al. 2001; S05). The presence of a dense gas component
(M$_{\rm gas} \sim 10^{8}$ M$_{\odot}$ with a high central density of 0.1 -
0.3 cm$^{-3}$) potentially provides the fuel for the central
supermassive black hole (SMBH), in environments where the amount of the
galactic cold gas is at a minimum. Moreover, their properties shed light
on the evolution history of their hosts.

For the previous galaxy coronae discovered in rich clusters, there is no
strong evidence for a high velocity of the host galaxy relative to the surrounding
ICM or for a strong nuclear activity of the host galaxy. In this paper, we show an extreme
example for which the galaxy is known to be moving very fast ($>$ 2500 km/s)
through a hot cluster and to host a powerful AGN with strong radio emission.
NGC~1265 (3C~83.1B, 4C~41.06) is the prototype of narrow-angled tail (NAT)
radio sources (Ryle \& Windram 1968). NGC~1265 is in the Perseus cluster,
27$'$ northwest of the cluster center (Fig. 1). Its radial velocity (7536 km/s,
Huchra, Vogeley \& Geller 1999) is $\sim$ 2300 km/s higher than that of the
Perseus cD galaxy NGC~1275 (5264 km/s, Huchra et al. 1999), indicating a large
infalling velocity for NGC~1265. Highly complicated radio morphologies
are revealed from 2 cm to 92 cm (Owen, Burns \& Rudnick 1978; O'Dea \& Owen
1986; Sijbring \& de Bruyn 1998).
Two narrow east-west radio jets emerge from the nucleus, but are quickly bent
towards the north to form two tails, which are further merged 4$'$ north
of the nucleus. Radio images at 49 cm and 92 cm (Sijbring \& de Bruyn 1998)
further reveal a low surface brightness extension of the tail that bends
at least three times over an angle of almost 360\mydegree .
The projected total length of the tail is at least 40$'$ (or 0.88 Mpc).
This complex morphology was tentatively explained as a result of the galaxy's
infalling orbit and the bulk flow of the ICM gas to the east (Sijbring \&
de Bruyn 1998). Despite the unique appearance of the galaxy in the radio,
optically it is a rather normal elliptical. A small dust lane is
revealed by the \hst\ image (Fig. 1), which is almost perpendicular to the
jet direction and aligned with the galaxy major axis. The galaxy is not a
strong X-ray emitter from the \rosat\ data (Rhee, Burns \& Kowalski 1994).
We obtained a deep (94 ksec) \chandra\ observation on this galaxy
and the X-ray source is fully resolved.

The velocity of the Perseus cluster is 5366 km/s (Struble \&
Rood 1999). We use a redshift of 0.018, to calculate the luminosity distance
of NGC~1265, assuming H$_{0}$ = 70 km s$^{-1}$
Mpc$^{-1}$, $\Omega$$_{\rm M}$=0.3, and $\Omega_{\rm \Lambda}$=0.7.
The luminosity distance is 78.4 Mpc and 1$''$ corresponds to
0.366 kpc. We use the Galactic absorption of 1.46$\times$10$^{21}$
cm$^{-2}$, which is consistent with the absorption column derived from
the X-ray analysis. The solar photospheric abundance table by Anders \& Grevesse
(1989) is used in the spectral fits. Uncertainties quoted are 1 $\sigma$.

\section{\chandra\ observation}

\subsection{\chandra\ data reduction}

A 93.9 ksec \chandra\ observation was performed with the Advanced CCD Imaging
Spectrometer (ACIS) on March 15-16, 2003. The optical axis lies on the CCD S3 and the data
were telemetered in Very Faint mode. Standard data analysis is performed
which includes the correction for the slow gain
change\footnote{http://cxc.harvard.edu/contrib/alexey/tgain/tgain.html}.
Since the CCD S1 is off, we investigated the background light curve
from the source-free region on the CCD S3. Excluding time intervals with
significant background flares results in a total exposure of 65.7 ksec.
Although the cluster emission is still significant across the whole
\chandra\ field, any weak background flares do not affect our analysis
on small scales, since a local background is used.
For the analysis of the cluster diffuse emission, we used the period
D background file, ``aciss\_D\_7\_bg\_evt\_271103.fits''
\footnote{http://cxc.harvard.edu/contrib/maxim/bg/index.html}.
The particle background level (measured in PHA channels 2500-3000 ADU)
was 1.1\% higher than that of the period D background data. Thus, we
increased the background normalization by 1.1\%.

We corrected for the ACIS low energy quantum efficiency (QE)
degradation, which increases with time and is positionally dependent
\footnote{http://cxc.harvard.edu/cal/Acis/Cal\_prods/qeDeg/index.html}.
The calibration files used correspond to \chandra\ Calibration Database
3.0.0 from the \chandra\ X-ray Center. In the spectral analysis, a
low energy limit of 0.5 keV is used to minimize the calibration
uncertainties in low energy.

\subsection{The spatial properties of the NGC~1265 X-ray source}

Despite the gigantic extent ($> 40'$) of NGC~1265 in the radio (O'Dea \& Owen
1986; Sijbring \& de Bruyn 1998), its X-ray extent is tiny ($\sim 10''$),
as shown in Fig. 1. The X-ray emission is composed of a central point source
and diffuse emission. The central point source is located within 0.1$''$ of
the nucleus determined from the \hst\ observation (Fig. 1b) \footnote{The
\chandra\ astrometry is generally better than 0.5$''$ (http://cxc.harvard.edu/cal/ASPECT/celmon/)}. 
The soft X-ray
emission (0.5 - 1.5 keV) is asymmetric,
with a sharp edge 3$''$ south of the nucleus and an extension to the north.
The asymmetry is quantitatively shown by the 0.5 - 1.5 keV surface brightness
profiles (exposure-corrected) along the south, north and east-west sectors (Fig. 2).
The sectors are defined in the caption of Fig. 2.
The 0.5 - 1.5 keV surface brightness profiles in the east and west differ
very little so we combine them. The surface brightness to the south decreases
rapidly at $\sim 2.5''$ and reaches the background level beyond $\sim 3''$,
while the brightness profile to the north show excess above the local background
to at least 25$''$. This morphology implies the motion of the X-ray source 
towards the south and the action of ram pressure by the surrounding ICM.
The azimuthally averaged profile is flat beyond 30$''$, which marks the
local background level.

The nuclear point source dominates the hard X-ray emission (Fig. 1a). As there
is no statistically important difference on the 2 - 6 keV surface brightness
profiles in the south and north, only the azimuthally averaged profile is shown (Fig. 2).
The point spread function (PSF) of the central point source was derived
with the \chandra\ Ray Tracer (ChaRT), assuming the best-fit absorbed powerlaw
spectrum derived from the spectral analysis (discussed later in $\S$2.3).
Since the X-ray source is only 7.1$''$ from the optical axis of the observation,
the PSF is symmetric. In the 0.5 - 1.5 keV band, the
PSF size is much smaller than the source size (Fig. 2). In the 2 - 6 keV
band, a model with the PSF plus the local background matches the data
well within the inner $\sim 1.6''$, but under-estimates the observed
brightness of 2$'' - 8''$ at a level of 5.2 $\sigma$.
This hard X-ray excess is indicative of low mass X-ray binary (LMXB)
emission in NGC~1265, which will be discussed next.

We assume that the LMXB light profile follows the stellar light profile
which is derived from the \hst\ observation. A bright star 3$''$ east
of the nucleus distorts the galaxy image and affects
the ground photometry of the galaxy. The center of the star is saturated
in the HST F702W image, but not in the F673N image. We subtract the star
light and derive the light profile of the galaxy. The galaxy contributes
70\% of the total light in the F702W image. The stellar type is unknown, so we
simply adopt the same percentage for the B band. The total B magnitude of
the galaxy measured from the ground is 12.45 mag after extinction correction
and K-correction. Thus, L$_{\rm B} = 6.9 \times 10^{10}$ L$_{\odot}$ for
NGC~1265. From the L$_{\rm LMXB}$ - L$_{\rm B}$ scaling relation derived by
Sarazin, Irwin \& Bregman (2001), the expected LMXB light distributions
in the 0.5 - 1.5 keV and 2 - 6 keV band are shown in Fig. 2. In the hard
band, a model with a central AGN (PSF), local background and LMXB light
roughly re-produces the data, but over-estimate the observed emission by
$\sim$ 15\%. In the soft band, the LMXB emission is able to account for
the emission beyond 9$''$, where the surface brightness profile is much
flatter than that within 9$''$. However, as the LMXB light does not
blur the southern sharp edge (Fig. 2), the LMXB emission must be discrete
and dominated by those brightest binaries. In the following, we
constrain the spectral properties of the X-ray source and the
gas properties of the corona.

\subsection{The spectral properties of the NGC~1265 X-ray source}

The spectral properties of the NGC~1265 X-ray source were measured.
In view of the asymmetry, we extracted the global spectrum from a 10$''$
radius circle centered 4.6$''$ north of the nucleus. The region is chosen
to enclose all X-ray emission of NGC~1265 above the local background (Fig. 2),
although the X-ray bright region is considerably smaller (Fig. 1). The
background
spectrum was extracted from the 17$'' - 73''$ annulus centered on the
nucleus. We first used a model with two components (POWERLAW + MEKAL).
If only Galactic absorption is applied, the fit is acceptable but the
nuclear source has an unusually flat spectrum (Table 1). The presence
of a dust lane on the \hst\ image may imply extra absorption around the
central source. Thus, we allowed additional absorption for the
POWERLAW component. As shown in Table 1, the fit is much improved
and the derived powerlaw index ($\sim$ 2.0) is typical for an AGN. An absorbed nucleus
is also supported by the flat 0.5 - 1.5 keV surface brightness
profile within the central 1$''$ (Fig. 2). In any case, the corona
temperature measurement is robust (0.60 - 0.65 keV). However, in this
fit, the LMXB emission (not suffering the excess absorption column)
cannot be separated from the nuclear emission. To better
separate the thermal corona, the central AGN and the LMXB emission,
we extracted spectra in two regions, within 1.6$''$ radius of the
nucleus (including nearly 90\% of the nuclear emission, Fig. 2), and the
global region excluding the central 1.6$''$ (1.6$'' - 10''$, but
note the different centers). Separate spectral fits were performed as
shown in Table 1. For the inner region, we apply a model with a MEKAL
plus an absorbed POWERLAW. For the outer region, we apply a model with a
MEKAL plus thermal bremsstrahlung emission from LMXB. The background-subtracted
spectra are shown in Fig. 3 with the best fits of the two-component model.
Abundances cannot be well constrained (particularly the upper bound) and
are fixed at the solar value. The spectral fits to both regions give an
abundance lower limit of 0.5 solar (90\% confidence level). The allowed
abundance change and the inclusion of the hard component do not affect the
robustness of the temperature measurement.

We derived the deprojected temperature of the inner region (the fifth
row in Table 1) with the standard non-parametric ``onion peeling'' technique).
The spectral properties of the 1.6$'' - 3.2''$ annulus,
where the X-ray emission is still rather symmetric, were also examined
(the sixth row in Table 1). Gas temperature decreases from 0.7 keV at the
outskirts to 0.45 keV within the central 0.6 kpc high-density core, implying
the action of radiative cooling. As shown in Fig. 2, the central point
source also accounts for some emission in the outer region ($\sim$ 29\%)
and there may be LMXB emission within 1.6$''$ at the level of $\sim$ 5\%.
However, these extra components bring little change to the fits and
computed parameters (including luminosities). The best-fit absorption
excess for the AGN is $\sim 2\times$10$^{22}$ cm$^{-2}$.
The 2 - 10 keV and bolometric luminosities of the central AGN are
5.5$\times$10$^{40}$ ergs s$^{-1}$ and 3.0$\times$10$^{41}$ ergs s$^{-1}$
respectively. The 0.5 - 2 keV and bolometric luminosities of the corona
are 3.3$\times$10$^{40}$ ergs s$^{-1}$ and 4.8$\times$10$^{40}$
ergs s$^{-1}$ respectively. The 0.3 - 10 keV luminosity of the
1.6$'' - 10''$ LMXB component is 1.8$\times$10$^{40}$ ergs s$^{-1}$.

We also examined the spectral properties of the north extension.
Its spectrum was extracted from the region defined in Table 1. The result
confirms that the north extension is from the emission of thermal gas,
while the LMXB component only contributes to $\sim$ 5\% of the 0.5 -
1.5 keV counts there. The soft coronal emission dominates
in the 0.5 - 1.5 keV band. Within 1.6$''$ radius, the nuclear emission
only contributes to 6\% of the total 0.5 - 1.5 keV brightness. Although
the LMXB emission may contribute to $\sim$ 24\% of the total 0.5 - 1.5
keV emission in the 1.6$'' - 10''$ region, the contribution drops to 10\%
in a smaller region enclosing all soft X-ray bright region (a 6.5$''$ radius circle
centered at 3.2$''$ north of the nucleus). Thus, the 0.5 - 1.5 keV surface
brightness well delineates the morphology of the soft corona.

\subsection{The properties of the surrounding ICM}

We need to know the properties of the surrounding ICM to study the corona-ICM
interaction. If NGC~1265 is on the
plane of the sky (27$'$ from the cluster center), the electron density of the
surrounding ICM is 5.9 $\times$ 10$^{-4}$ cm$^{-3}$ from the \rosat\ data
(Ettori, Fabian \& White 1998). If the galaxy is within 20$'$ (or 439 kpc)
of the plane of the sky along the line of sight, the ICM density is $> 4.1
\times 10^{-4}$ cm$^{-3}$. We take a typical ICM electron density of 5
$\times 10^{-4}$ cm$^{-3}$ in this work, but keep the uncertainty in mind.
The spectrum of the surrounding ICM (16$'' - 90''$) is studied,
as well as that of the whole ICM emission on the S3 CCD. 
The surrounding ICM (projected) is 10 times hotter ($\sim$ 6.7 keV, Table 1)
than the small galaxy corona ($\sim$ 0.6 keV). The \chandra\ ICM temperature
is consistent with the \asca\ value for this region (Furusho et al. 2001).

\subsection{The gas distribution of the NGC~1265 corona}

We want to characterize the gas distribution of the corona in different directions
with the derived 0.5 - 1.5 keV surface brightness profiles. The contribution
of the nuclear emission to the soft band and the LMXB emission have to be
subtracted. The first component is derived from the PSF and the spectral fits in $\S$2.3,
while the LMXB component is estimated from the optical light. Both components
are small contributions to the total X-ray emission within 8$''$. The LMXB
component in the south sector is not subtracted, as the scaled LMXB light from
the optical light largely over-estimates the actual LMXB light beyond 3$''$ in
the south. Nevertheless, any LMXB contribution to the bright core within 2$''$
is small. Since the X-ray source is so small, the \chandra\ PSF has to
be considered in the fits. We applied a model composed of a $\beta$ model +
a constant background for profiles in the south, north and east-west, as well
as the azimuthally averaged profile. We also tried to put a truncation radius
to the $\beta$ model, especially for the southern profile. The results are shown
in Table 2. Without a truncation radius, the fit to the southern profile yields
an unphysical $\beta$ ($\gsim$ 10), which is
thus not shown. The fits to the profiles in the south and north are shown in Fig. 4.
The southern profile is clearly truncated at the sharp edge (2.2$\pm0.3''$ south
the nucleus) with a large jump of surface brightness across the edge ($\sim$ 106),
while the truncations in other directions are not significant with a small surface
brightness jump of $<$ 3 - 5. Despite the large uncertainties, gas distributions
(assuming the same emissivity) in all directions are consistent with the same
one, but with different truncation radii (or no truncation). 
We also constrain the width of southern edge by smearing the truncated $\beta$ model
with a Gaussian exp(-r$^{2}$/2$\sigma^{2}$). The best-fit model requires $\sigma$ = 0
and the 95\% upper limit of $\sigma$ is only 0.2 kpc. We can compare the
NGC~1265 corona with the two large coronae in A1367 (S05). Both luminous coronae in
S05 are largely symmetric and the surface brightness jump across the presumably
pressure-confined boundary is at most 4 - 9. Thus, the sharp edge only 2.2$''$ (or
0.81 kpc) south of the NGC~1265's nucleus is certainly unique and cannot be
explained by the confinement of the still ICM pressure.

The gas density and mass can be derived from the surface brightness
distribution.
The best-fit thermal models of inner and outer regions give almost
identical emissivity, if the metallicity is chosen the same. For
simplicity, we assume a constant metallicity and a constant emissivity
for the whole corona. In view of the unsymmetric shape of the NGC~1265
corona, we applied two methods to derive the density profile and the
total gas mass. The first is to use the fit to the azimuthally averaged
0.5 - 1.5 keV profile derived above. Assuming a metallicity of 0.5 - 1.5
solar, the central electron density is 0.28 - 0.48 cm$^{-3}$, and the total
gas mass is 3.2 - 5.4 $\times 10^{7}$ M$_{\odot}$ (within r$_{\rm cut}$ of
7.7 kpc). The second method is to assume that the corona is composed of two
hemispheres characterized by the surface brightness profiles of the south
and the north respectively. Assuming a metallicity of 0.5 - 1.5 solar,
the central electron density is 0.22 - 0.37 cm$^{-3}$, and the total gas
mass is 2.4 - 4.2 $\times10^{7}$ M$_{\odot}$ (within r$_{\rm cut}$).
Thus, the central electron density of the corona is 0.22 - 0.48
cm$^{-3}$ and the total gas mass is 2.4 - 5.4 $\times10^{7}$ M$_{\odot}$.
Assuming a central abundance of 0.5 - 1.5 solar, the central gas cooling time
is only 5 - 15 Myr.

With the derived gas density profile, the X-ray ISM - ICM thermal pressure
ratios at the southern edge and the northern tail can be estimated.
We assume an ISM temperature of 0.69$\pm$0.04 keV (Table 1) and applied
Monte-carlo simulations to calculate the errors. The derived X-ray ISM - ICM
pressure ratios are 24$\pm$16 at the southern edge and 1.8$^{+1.4}_{-1.6}$
at the end of the northern tail. Although the errors are large, the values indicate
that the static ICM pressure is not enough to produce the southern sharp edge.

\subsection{X-ray emission from the jets}

X-ray emission, at the significance levels of 2.2 - 3.6 $\sigma$, is found
at the positions of three radio inner knots in the jets (E1, E2 and W1 from
O'Dea \& Owen 1986). A total of $\sim$ 28 counts are collected from these
three knots. Assuming a power law spectrum with a photon index of 2.0, the
0.5 - 10 keV luminosities are $\sim 1.1\times10^{39}$ ergs s$^{-1}$,
6$\times10^{38}$ ergs s$^{-1}$ and 5$\times10^{38}$ ergs s$^{-1}$ for E1,
E2 and W1 respectively. Recent \chandra\ observations show that X-ray jet
is common in galaxies with radio jets (e.g., Worrall, Birkinshaw \& Hardcastle
2001; Sambruna et al. 2004). However, this deep observation of NGC~1265
implies that X-ray jet can be very faint in some cases (at least 40 times
fainter than the faintest jet in Worrall et al. 2001). We estimate the
1 keV X-ray to 8-GHz radio flux density ratio over the X-ray detected
regions of the jet, $\sim 10^{-8}$, which is smaller than what is generally found
for low-power radio galaxies where the jet X-ray emission is considered
as synchrotron emission ($\sim 10^{-7}$, see the list in Evans et al. 2005).
Non-detection of NGC~1265's optical jet is also against the synchrotron
interpretation.

In spite of the non-detection of the continous X-ray jets, there are spatial
indentations east and west of the corona (Fig. 1a) that are aligned with the
radio jets. These features imply the interaction between the narrow jets
and the X-ray corona. The widths of the indentations are comparable to the
widths of the radio jets.

\section{Discussion}

\subsection{The X-ray edge \& the velocity of NGC~1265}

The sharp X-ray edge is a strong indication of the ISM-ICM interaction.
The mean free paths of particles near the edge are: $\lambda_{\rm ICM}$ =
27 kpc, $\lambda_{\rm ISM}$ = 1.5 (n$_{\rm e, ISM}$/0.1 cm$^{-3}$)$^{-1}$ pc,
$\lambda_{\rm ICM \rightarrow ISM}$ = 60 ($\lambda_{\rm ISM}$/1.5 pc) pc,
and $\lambda_{\rm ISM \rightarrow ICM}$ = 5.4 kpc, assuming temperatures
of 6.7 keV and 0.7 keV for the ICM and the ISM near the edge respectively.
As $\lambda_{\rm ICM}$ is much larger than the size of the corona ($\sim$
4 kpc in diameter), the ICM may not be treated as fluid relative
to the corona. Furthermore, as the galaxy with only stars is rather porous
and the corona is so small, it is questionable whether a bow shock
can be generated ahead of the corona to change the fluid pattern there.
A collisionless bow shock may still be generated as implied by the weak
92 cm emission detected ahead of the X-ray edge (Sijbring \& de Bruyn
1998), but this depends on the magnetic field structure around the corona.
The magneto-hydrodynamic (MHD) interaction between the galactic magnetic
field and the ICM particles (and the magnetic field frozen into them) is
however poorly understood. $\lambda_{\rm ISM \rightarrow ICM}$ derived above
is much larger than the width of the edge ($<$ 0.2 kpc), which can only be
understood if the actual mean free paths of particles are largely suppressed
by the magnetic field at the ICM-ISM boundary. Recent simulations of gas
stripping and the evolution of X-ray coronae in clusters (e.g.,
Toniazzo \& Schindler 2003; Acreman et al. 2003) tried to study the problem
with only fluid dynamics, but did not address the small sizes of X-ray coronae,
nor the inclusion of MHD. The observed properties of the NGC~1265 corona
demonstrate that it is necessary to include magnetic fields, especially
for small coronae moving in the hot clusters as the mean free path of
particles is sensitive to temperature ($\propto$ T$^{2}$/n).

Our knowledge of the magnetic field in elliptical galaxies is poor
(reviewed by Widrow 2002).
Moss \& Shukurov (1996) proposed two types of seed fields in elliptical galaxies:
stellar magnetic field ejected by SNe and stellar winds, and magnetic remnants
that arise if elliptical galaxy forms from mergers of spirals. These seed fields
have to be amplified by dynamos. They proposed two types of dynamos: acoustic
turbulence driven by SNIa and stellar winds, and vortical turbulence driven by
stellar motion. Since the magnetic Reynolds number is large, the amplified
magnetic field should be frozen in the gas (or the hot gas in elliptical galaxies).
When the galaxy falls into the dense ICM, the galactic magnetic field may
be compressed and stretched at the boundary, responsible for the suppression
of transport processes. However, this simple picture may be too naive as there
are a lot of details unclear. Moreover, the small corona of NGC~1265 is
surrounded by the hot ICM filling the NGC~1265 galaxy. How the field within the
corona can be magnetically isolated from the field in the surroundings (still
in the galaxy) is unclear, especially if the turbulent diffusion of magnetic
field is important (e.g., Lesch \& Bender 1990).

Assuming the radial velocity difference of NGC~1265 and the Perseus
cluster is NGC~1265's radial velocity relative to the cluster, the velocity
of NGC~1265 (v$_{\rm gal}$) is 2170/cos$\theta$ km/s, where $\theta$
(between 0 and 90 degree) is the angle between the line of sight and
the opposite of the moving direction. The radial velocity dispersion of
the Perseus cluster is 1324 km/s (Struble \& Rood 1999), which corresponds
to a 3D velocity dispersion of 2293 km/s ($\sigma$). NGC~1265's velocity
cannot be too high (e.g., 3$\sigma$, or 6880 km/s, $\theta$ = 71\mydegree.6)
for the galaxy to remain bound.

The sharpness of the X-ray edge implies that $\theta$ is not small.
Otherwise, the edge will be smeared by projection.
In principle, we can use the sharp X-ray edge to constrain
the velocity of the galaxy. This first requires a good understanding of
projection. We performed a number of simulations. The shape of the corona is
assumed to be prolate with the nucleus at the center. Since the front
of the corona is ablated and compressed by ram pressure, the front
is approximated by a large sphere cut by the prolate ellipsoid, while the center
of the sphere lies north of the nucleus. The small corona is assumed to be
axially symmetric to the direction of the motion. The coronal density profile
is represented by a $\beta$ model with a truncation at the boundary of
the corona (see $\S$2.5). The projected morphology of the simulated corona matches
the observed morphology well. We compare the projected profiles with the observed ones
and find that the X-ray sharp edge cannot be preserved for $\theta <$ 45\mydegree .
As $\theta$ decreases from 90\mydegree , the front edge has
to be shifted towards the nucleus to allow the projected edge to match the observation.
Thus, the best-fit velocity increases as $\theta$ decreases,
opposite from the dependence of velocity on $\theta$ from the
radial velocity constraint. Therefore, the velocity can be in principle
constrained well by these two methods. However, several uncertainties
prevent a tight constraint on the velocity. First, there is a
degeneracy between the ICM density and the galaxy velocity. Second,
the coronal density can vary 70\% for an abundance change of 0.5 - 1.5 solar.
Lastly, the MHD properties of the ICM are not known.
For an ambient density of 5$\times10^{-4}$ cm$^{-3}$, the best-fit
angle is 50\mydegree\ - 65\mydegree\ and the velocity is 3500 - 5000 km/s
if we treat the ICM as fluid.
Two interesting facts are: 1) the surrounding ICM density is
$> 2\times10^{-4}$ cm$^{-3}$ if NGC~1265's velocity is $<$ 5000 km/s
and the abundance of the corona is $<$ 1.2 solar; 2) the abundance of
the corona has to be $<$ 3 solar (a tighter upper limit than that derived
from the spectral analysis) for the ISM gas to be dense enough
to withstand the ram pressure.
Combining these constraints, we conclude (conservatively):
45\mydegree\ $< \theta <$ 67\mydegree\ and 3100 km/s $<$ v$_{\rm gal} <$
5500 km/s (note the local sound speed is 1.3$\times10^{3}$ km/s).

\subsection{The evolution of the NGC~1265 corona}

As the corona moves through the ICM with v$_{\rm gal}$ of $>$ 3100 km/s, it
is subject to ram pressure
stripping and other stripping processes by the surrounding ICM (e.g.,
Nulsen 1982). The analysis of ram pressure stripping in S05 ignores the
high ISM pressure. The survival of NGC~1265's corona in the face of high
speed ram pressure stripping and the presence of a sharp edge indicate that
high thermal pressure of the inner cooling core is the key to preserve
the corona from ram pressure stripping, although the outskirts of the corona
are very vulnerable to stripping.

As was discussed in the last section, the physics around the small corona
should be best described by MHD. The Kelvin-Helmholtz (K-H) instability may
or may not be suppressed. Lacking a good understanding of the relevant
physics, only a qualitative discussion is presented. No matter how we
treat the surrounding ICM, as fluid or as particles, the typical mass-loss
rate of the corona by K-H instability or by the bombardment of the ICM
particles is approximately (e.g., Nulsen 1982):

\begin{eqnarray}
\dot{M}_{\rm strip} &\approx& \pi r^2 \rho_{\rm ICM} {\rm v_{gal}} \nonumber \\
&=& 0.64 (\frac{n_{\rm e, ICM}}{5 \times 10^{-4} {\rm cm}^{-3}}) (\frac{r}{2 {\rm kpc}})^2
(\frac{{\rm v}_{\rm gal}}{3500 {\rm km/s}}) {\rm M_{\odot} / yr}
\end{eqnarray}

This process acts at the boundary of the corona. We calculate the stellar
mass injection rate in these two regions, $< 1.6''$ and 1.6$'' - 10''$.
Using the generally adopted mass injection rate $\dot{M}_*$ = 0.15
M$_{\odot}$ yr $^{-1}10^{10}$ L$_\odot^{-1}$ (Faber \& Gallagher 1976)
and the optical luminosity in the chosen region (from the \hst\ data),
we estimate 0.04 M$_{\odot}$/yr and 0.15 M$_{\odot}$/yr for the inner
and outer region respectively. The mass deposition rate by cooling,
$\dot{M} \approx 2\mu$m$_{\rm p}$L$_{\rm X, bol}$/5kT, is calculated in
each region to be 0.19 M$_{\odot}$/yr and 0.16 M$_{\odot}$/yr. Thus, in the inner
region, the mass deposition rate is higher than the expected stellar
mass injection rate, while in the outer region, the expected mass loss rate by
stripping is comparable (if not higher, because of the uncertainties of
equation 1) to the expected stellar mass injection rate.
Although the mass loss by stripping decreases rapidly to be $\lsim$ the
stellar injection at radius $<$ 1 - 1.6$''$, the cooling rate changes little, if
there is no feedback on this small scale. The estimated life
time of the current corona is then only 0.1 - 0.2 Gyr.

There are several factors that can reduce the effects of stripping and offset cooling
to help the corona to survive much longer. First, the stripping by transport
processes can be suppressed by magnetic field. Second,
stellar mass injection is a key (e.g., shown in the simulations by Acreman
et al. 2003). While the Faber \& Gallagher (1976) result is an average, the
host galaxies with surviving coronae may have systematically larger stellar
mass injection rates than the average. 
Stellar mass injection outside the corona can also affect the ICM flow pattern
there, although the injected ISM gas is not likely to sink into the corona
before being stripped and mixed with the ICM. For NGC~1265, if we compare
the mass flow rate of the ICM and the expected stellar mass injection rate
outside the corona upstream, on the cross-section of the corona, the latter
is only $\sim$ 10\% of the ICM mass flux, and is not likely to be a significant
factor. However, combined with the galactic potential, magnetic field and
stellar motion, the ICM flow inside the galaxy (but outside the corona) should
be much more complicated than the free flow far away from the galaxy. This may
significantly affect the stripping rate estimated by Equation 1. 
Third, SNIa in principle can provide enough heating energy. Using the SN
rate estimates from Cappellaro, Evans \& Turatto (1999), the SNIa rate within
the size of the NGC~1265 corona is $\sim$ 0.2 / 100 yr. Assuming an energy
release of 6 $\times 10^{50}$
ergs per SNIa event, the total energy release of SNe Ia is 3.8 $\times 10^{40}$
ergs s$^{-1}$, which is comparable to the X-ray luminosity of the corona.
However, it has been long known that most of SN energy does not couple to
the hot gas from the observed L$_{\rm X}$ - L$_{\rm B}$ relation of the X-ray coronae
of early-type galaxies (e.g., Brown \& Bregman 1998). Thus, the question is
how SNIa dissipates its kinetic energy into the hot gas. Lastly, Vikhlinin et al.
(2001) proposed that largely reduced heat flux from the surrounding ICM can offset
cooling inside the corona if the heat conductivity in the corona is close
to the Spitzer value.

As shown in Vikhlinin et al. (2001) and S05, the large temperature gradient
from the ICM to the ISM implies that heat conduction must be suppressed.
Assuming a corona radius of 2 kpc and an ambient density of 5$\times10^{-4}$
cm$^{-3}$, we estimate for NGC~1265 that heat conduction has to be suppressed
by at least a factor of $\sim$ 60 at the ICM-ISM boundary, using the method presented in S05.

We notice that NGC~1265's ISM must be multi-phase inside the small corona, as
implied by the presence of the dust lane. Dust with mass $\lsim 10^{4}-10^{5}$
M$_{\odot}$ was detected in $\sim$ 80\% of nearby large elliptical galaxies
(e.g., van Dokkum \& Franx 1995). Dust detection in radio galaxies is $\sim$ 2
times higher than in radio-quiet galaxies, and the dust lane in radio galaxies
is usually perpendicular to the radio axis (van Dokkum \& Franx 1995), just as
is found in NGC~1265. It has been suggested that the central dust in
elliptical galaxies is the product of stellar mass loss from red giant
stars (Mathews \& Brighenti 2003). This scenario links dust with the
10$^{7}$ K coronal gas (also thought to have a stellar origin), although it
is unknown what the cooling product of the hot gas is. Both cold
dust and the cooling product of the high-density X-ray gas (related or not)
can provide fuel to the SMBH.

\subsection{The central AGN and the jets}

NGC~1265's nucleus is detected in both radio and X-ray (Fig. 1a).
The radio observations show that the central radio source is a compact
flat-spectrum source ($<$ 0.1$''$, Owen et al. 1978), with a
total radio luminosity of 7.3$\times10^{38}$ ergs s$^{-1}$ (10 MHz - 5 GHz).
The X-ray nuclear source is absorbed (N$_{\rm H}$ = 1.5-3$\times$10$^{22}$
cm$^{-2}$) with a 2 - 10 keV luminosity of $\sim 5.5\times$10$^{40}$
ergs s$^{-1}$. This high amount of absorption is unusual for FR I radio
galaxies, as only NGC~4261 (3C~270) has a similar high absorption column
in a systematic study of 25 FR I radio galaxies (Donato, Sambruna \& Gliozzi
2004). The conclusion of Donato et al. (2004) is that FR I radio galaxies
generally lack a standard optically thick torus, but the exception of
NGC~1265 implies that its nuclear X-ray emission comes from somewhere
close to the nucleus and some optically thick clouds exist between the
nucleus and our line of sight.

The central AGN can have significant impacts on the small corona through
its jets, especially in outbursts. NGC~1265 is luminous in radio (3C~83.1B),
8.9 Jy at 1.4 GHz. The radio flux index is 0.67 (from NASA/IPAC Extragalactic
Database). From 10 MHz to 5 GHz, the total radio luminosity is 3.7$\times10^{41}$
ergs s$^{-1}$, while the luminosity of the two jets is 2.6$\times10^{40}$
ergs s$^{-1}$ (O'Dea \& Owen 1987, distance adjusted to our value).
The jet power is known to be much larger than the radio luminosity of the
source. The jet power is distributed in relativistic electrons, ions and
magnetic field. The power in ions usually dominates. As the jets of NGC~1265
are bent by the ICM ram pressure, the kinetic power of its jets can be estimated
(e.g., Jones \& Owen 1979). We assume that the jet flow is non-relativistic,
while the power of relativistic jets is much larger.
From Euler's equation, we have:
$\rho_{\rm j}$ v$_{\rm j}^{2} \approx$ P$_{\rm ram}$ R / r$_{\rm j}$, where
$\rho_{\rm j}$ is the particle density in the jet, v$_{\rm j}$ is the velocity
of the jet flow, r$_{\rm j}$ is the radius of the jet, R is the radius of curvature
of the bent jet, and P$_{\rm ram}$ is the ICM ram pressure.
The kinetic power of jets is: L$_{\rm K} \approx$ $\pi$ r$_{\rm j}^{2}$
$\rho_{\rm j}$ v$_{\rm j}^{3}$. With the relation from the bending of jets
and adopting r$_{\rm j}$ = 1 kpc, R = 12 kpc (from the radio image),
v$_{\rm j}$ = 10$^{4}$ km/s, n$_{\rm e, ICM}$ = 5$\times10^{-4}$ cm$^{-3}$
and v$_{\rm gal}$ = 3500 km/s, L$_{\rm K}$ = 4.3 $\times 10^{43}$ ergs s$^{-1}$.
r$_{\rm j}$ could be smaller but v$_{\rm gal}$ may be larger. Their product
should not change a lot based on the relation from the bending of jets.
The energy in magnetic field is only $\sim$ 10\% - 25\% of L$_{\rm K}$
if the minimum pressure magnetic field is adopted (20 - 30 $\mu$G from
O'Dea \& Owen 1987). The AGN has been active for $> 10^{8}$ yr from the
length of the tail. We estimate that the total jet power over the active
phase of the nucleus is at least 1600 times the thermal
energy in the current corona. In view of the great vulnerability of
the current corona to AGN feedback, we conclude that the jets carry
great amount of energy without dissipation through the small corona,
as we previously suggested in S05 for weaker radio sources associated with
the coronae in NGC~3842 and NGC~4874.

The interaction of jets and X-ray plasma is highly suggestive from the
small X-ray indentations 2.5$''$ east and west of the nucleus (Fig. 1). 
The radio emission of the jets is anti-correlated with the X-ray emission
of the corona (Fig. 1), as the radio emission of the jets turn on after they leave the dense corona.
The jets may undergo a transition after they leave the dense corona, possibly
because of the change in the external pressure. The coronal thermal pressure,
1.9$\times10^{-10}$ (n$_{\rm e}$/0.1 cm$^{-3}$) (kT/0.6 keV) dynes cm$^{-2}$,
is larger than the minimum pressure in the inner knots of the jets,
6-9$\times10^{-11}$ dynes cm$^{-2}$ (O'Dea \& Owen 1987), while the
ICM thermal pressure is only 10$^{-11}$ dynes cm$^{-2}$. Thus, the ISM
thermal pressure may be enough to confine the jets, while the ICM pressure
may not. The edge-brightening of the jets from
the first knots implies the action of ram-pressure (O'Dea \& Owen 1986).
The jets are clearly bent by the ICM ram pressure, rather than the
pressure gradient in the ISM, as originally suggested by Jones \& Owen (1979).

If the dense core of the corona cools, the cooling product may feed the central SMBH.
The stellar velocity dispersion of NGC~1265 is unknown. Using the Faber-Jackson relation
and M$_{\rm BH}$-$\sigma$ relation (Tremaine et al. 2002), the mass of the SMBH at the
nucleus of NGC~1265 is $\sim 5\times10^{8}$ M$_{\odot}$. Assuming this
mass and a central gas temperature of 0.45 keV, the Bondi accretion radius
of the central SMBH is 36 pc (or 0.1$''$) if the surrounding coronal gas has
little residual bulk motion relative to the SMBH. The Bondi accretion rate
and the accretion luminosity are 0.0031 M$_{\odot}$/yr and 1.8$\times10^{43}$
ergs s$^{-1}$ (assuming a radiation efficiency of 0.1) for a central electron
density of 0.3 cm$^{-3}$. The Bondi accretion luminosity is known to largely
over-estimate the luminosity of many low luminosity AGN (e.g., Loewenstein
et al. 2001). The possible explanations include that the accretion process is
radiation inefficient, and other processes (e.g., feedback) reduce
the Bondi accretion rate significantly (e.g., Gliozzi, Sambruna \& Brandt 2003).

\section{Conclusion}

The main results of this deep observation of the prototype
NAT galaxy NGC~1265 are:

1. A small ($\sim$ 4 kpc) and asymmetric thermal corona is detected
around the nuclear region of NGC~1265. A sharp edge is found 2.2$''$
(0.8 kpc) south of the nucleus, while a $\gsim 8''$ extension is
present to the north. This indicates a high velocity motion of NGC~1265
towards the south on the plane of sky, which is consistent with the
long radio tail to the north.

2. The southern edge is very sharp, with a jump of surface brightness of
$\sim$ 106 and a width of $<$ 0.2 kpc. As the mean free paths of particles
in the ICM (27 kpc) and from the corona to the ICM (5.4 kpc) are much
larger than the width of the edge, magnetic field around the edge is
required to significantly reduce the particle diffusion. The interaction
between the small coronae and the ICM must be studied by MHD.

3. The temperature of the coronal gas decreases from 0.7 keV at the
outskirts to 0.45 keV at the center, while the temperature of the
surrounding Perseus ICM (projected) is $\sim$ 6.7 keV. The coronal abundance
is consistent with solar. The 0.5 - 2 keV luminosity of the corona is
3.3$\times10^{40}$ ergs s$^{-1}$.
The central electron density is $\sim$ 0.22 - 0.48 cm$^{-3}$, while the
total gas mass is $\sim 4\times10^{7}$ M$_{\odot}$.

4. An absorbed (N$_{\rm H}$ = 1.5 - 3 $\times10^{22}$ cm$^{-2}$) low luminosity
X-ray nucleus is detected, with a 2-10 keV luminosity of
$\sim 5.5\times10^{40}$. This amount of high nuclear absorption column is
not usual for FR I radio galaxies.

5. Three inner radio knots are detected in X-rays. Indentations east and
west of the corona indicate the interaction between jets and
hot gas. There is an anti-correlation between the emission of the
radio jets and the X-ray corona (Fig. 1).
This implies that jets have undergone a transition as they leave the
dense corona.

6. The great vulnerability of NGC~1265's corona to the large power carried
by jets from the central SMBH implies that AGN deposit all energy
outside the current corona. This also may imply that AGN heating cannot quench
the cooling very close to the nucleus (several kpc) in ``cooling-flow'' clusters.

7. Constraints from the radial velocity and the
X-ray edge yield a velocity for NGC~1265 between 3100 and
5500 km/s, and the angle between the tail and the line of sight is
45\mydegree - 67\mydegree .

8. If the stripping, cooling and evaporation are not suppressed, the
current corona can only survive for 0.1 - 0.2 Gyr.
However, enhanced stellar mass injection, suppressed stripping and
SNIa heating (if energy coupled to the hot gas) can significantly
increase the lifetime of the corona. For survival of NGC~1265's corona,
heat conduction has to be suppressed by at least a factor of $\sim$ 60 at the
ICM-ISM interface.

The survival of NGC~1265's corona from ram pressure stripping implies
that coronae of similar or more massive galaxies can survive $\gsim$
1000 km/s stripping in environments like the core of the Coma cluster.
A systematic study based on \chandra\ data is underway, which will better
understand the evolution of these small corona in rich clusters, including
their fates, fueling of the SMBH, interaction with the environment, and
the relation with large cooling cores.

\acknowledgments

We thank Alexey Vikhlinin, Bill Forman \& Paul Nulsen for inspiring
discussions. We are grateful to C. P. O'Dea and Ylva Pihlstrom for providing us
the VLA images of NGC~1265 shown in Fig. 1. We thank the referee for helpful comments.
We acknowledge support from the NASA contracts G02-3184X and Smithsonian Institution.

\begin{table}
\begin{center}
\begin{small}
\caption{Spectral fits of the NGC~1265 X-ray source}
\vspace{0.2cm}
\begin{tabular}{rccccc}
\hline \hline
Region & \multicolumn{2}{c}{Thermal gas} & \multicolumn{2}{c}{AGN} & \\
 & T (keV) & Z ($\odot$) & N$_{\rm H}$ (10$^{22}$ cm$^{-2}$)$^{\rm a}$ & $\Gamma$ & $\chi^{2}$/d.o.f. \\ \hline
$<$10$''$ (total)$^{\rm b}$ & 0.62$^{+0.02}_{-0.03}$ & (1.0) & - & 0.94$\pm$0.12 & 76.9/75 \\
               & 0.63$^{+0.02}_{-0.03}$ & 1.2$^{+4.9}_{-0.6}$ & 1.3$\pm$0.3 & (1.7) & 64.1/74 \\
               & 0.63$^{+0.03}_{-0.02}$ & 0.9$^{+4.0}_{-0.2}$ & 1.8$^{+0.9}_{-0.6}$ & 2.04$^{+0.32}_{-0.15}$ & 63.3/73 \\ 
$<$1.6$''$ $^{\rm c}$ & 0.51$\pm$0.04 & (1.0) & 2.4$^{+0.3}_{-0.4}$ & 1.97$^{+0.27}_{-0.20}$ & 34.7/42 \\
$<$1.6$''$ (deproj.)$^{\rm c}$ & 0.45$^{+0.05}_{-0.03}$ & (1.0) & 2.4$^{+0.5}_{-0.7}$ & 1.98$^{+0.13}_{-0.20}$ & 36.1/42 \\
1.6$'' - 3.2''$ $^{\rm d}$ & 0.63$^{+0.05}_{-0.04}$ & (1.0) & - & - & 23.7/31 \\
1.6$'' - 10''$ $^{\rm e}$ & 0.69$\pm$0.04 & (1.0) & - & - & 36.4/41 \\
North extension$^{\rm f}$ & 0.76$^{+0.07}_{-0.08}$ & (1.0) & - & - & 8.2/16 \\
16$'' - 90''$ $^{\rm g}$ & 6.67$^{+0.68}_{-0.70}$ & 0.41$^{+0.41}_{-0.32}$ & - & - & 102.6/113 \\
S3 CCD$^{\rm h}$ & 6.41$\pm$0.25 & 0.48$^{+0.10}_{-0.09}$ & - & - & 192.0/143 \\
\hline\hline
\end{tabular}
\begin{flushleft}
\leftskip 1pt
$^{\rm a}$ The excess absorption column (besides the Galactic value of 1.46$\times10^{21}$
cm$^{-2}$) for the nuclear source \\
$^{\rm b}$ The global spectrum of the NGC~1265 X-ray source, while the center is shifted to 4.6$''$ north of the nucleus to account for the asymmetric morphology of the corona. The background is from the surroundings (all the same for corona emission; see text).\\
$^{\rm c}$ Centered on the nucleus \\
$^{\rm d}$ The annulus centered on the nucleus. A 8 keV Bremsstrahlung component is included to model the contribution from LMXB.\\
$^{\rm e}$ The global region excluding the central 1.6$''$ (radius) region. A 8 keV Bremsstrahlung component is included.\\
$^{\rm f}$ The region is defined as the north half of a 3.5$'' \times 8''$ (east-west semi-minor axis
$\times$ north-south semi-major axis) ellipse centered 1.6$''$ north of the nucleus, excluding the region
within 2.6$''$ radius of the nucleus. A 8 keV Bremsstrahlung component is included.\\
$^{\rm g}$ The surrounding ICM emission, centered on NGC~1265's nucleus, with the blank sky
background excluded \\
$^{\rm h}$ The integrated ICM emission on the S3 CCD field \\
\end{flushleft}
\end{small}
\end{center}
\end{table}

\begin{table}
\begin{small}
\begin{center}
\caption{The fits of S$_{\rm X}$ in different directions}
\vspace{0.3cm}
\begin{tabular}{ccccccc}
\hline \hline
Direction & r$_{\rm c}$ & $\beta$ & r$_{\rm cut}$ & S$_{\rm X}$ jump & f$_{\rm back}$ & $\chi^{2}$/d.o.f. \\
       & (arcsec)       &         & (arcsec)      & (@r$_{\rm cut}$)  & (10$^{-6}$ cnts s$^{-1}$ arcsec$^{2}$) &  \\ \hline

South & 1.5$\pm$0.5 & 0.66$\pm$0.15 & 2.2$\pm$0.3 & $\sim$ 106 & 1.90$\pm$0.14 & 7.5/9 \\
North & 1.5$^{+0.5}_{-0.3}$ & 0.74$^{+0.09}_{-0.07}$ & - & - & 1.58$\pm$0.15 & 17.2/18 \\
 & 1.4$^{+0.5}_{-0.4}$ & 0.69$^{+0.10}_{-0.07}$ & 8.1$^{+1.8}_{-0.4}$ & $\sim$ 4.8 & 1.66$\pm$0.13 & 17.9/17 \\
East-west & 1.3$\pm$0.2 & 0.80$^{+0.09}_{-0.07}$ & - & - & 1.68$\pm$0.10 & 18.9/13 \\
 & 1.1$\pm$0.2 & 0.73$^{+0.08}_{-0.07}$ & 6.3$\pm$0.6 & $\sim$ 4.0 & 1.70$\pm$0.09 & 16.6/12 \\
All & 1.29$^{+0.29}_{-0.08}$ & 0.77$^{+0.04}_{-0.02}$ & - & - & 1.75$\pm$0.04 & 18.2/24 \\
 & 1.11$^{+0.18}_{-0.15}$ & 0.71$^{+0.03}_{-0.04}$ & 7.7$\pm$0.4 & $\sim$ 2.7 & 1.71$\pm$0.03 & 15.4/23 \\

\hline\hline
\end{tabular}
\vspace{0.2cm}
\begin{flushleft}
\leftskip 1pt
note: The $\beta$-model fits to the 0.5 - 1.5 keV S$_{\rm X}$ in different directions
(truncated or not). The local PSF is included in the fits. f$_{\rm back}$ is the flux of the local
background, which slightly decreases from the south (closer to the cluster center)
to the north.
\end{flushleft}
\end{center}
\end{small}
\end{table}

\begin{figure}
\vspace{-3.5cm}
  \centerline{\includegraphics[height=1.4\linewidth]{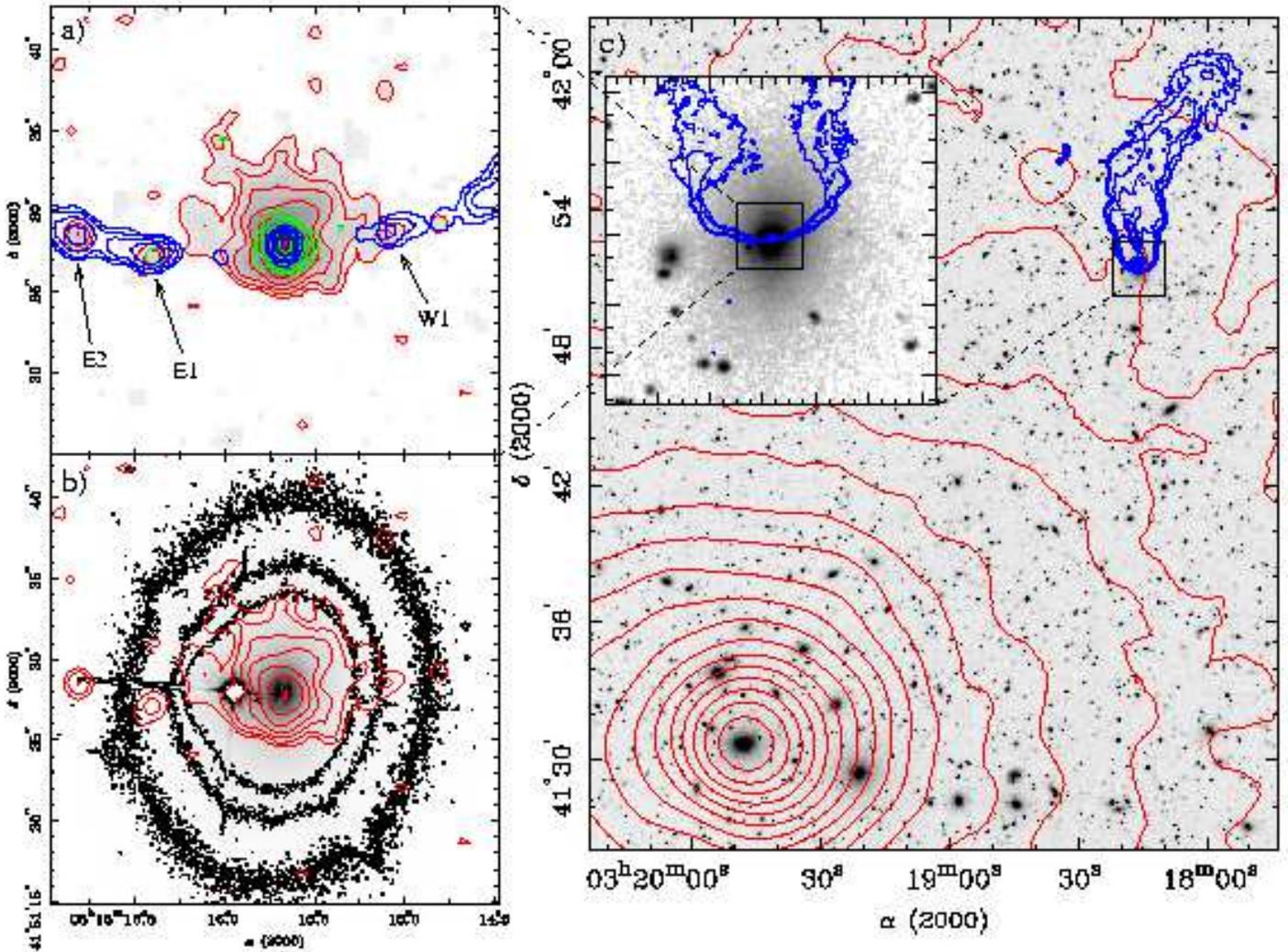}}
\vspace{-6cm}
  \caption{a): The 0.5 - 1.5 keV \chandra\ contours (red, from 0.023
to 1.5$\times10^{-3}$ cts s$^{-1}$ arcsec$^{-2}$, separated by a factor of 2)
superposed on its own image (exposure corrected, smoothed by one ACIS pixel).
The 2 - 6 keV \chandra\ contours (green, from 0.25 to 8$\times10^{-4}$
cts s$^{-1}$ arcsec$^{-2}$, separated by a factor of 2) are also shown
(exposure corrected, smoothed by one ACIS pixel). The sharp edge $\sim 3''$
south of the nucleus is prominent, as well as the extension to the north.
A 20 cm VLA image of the jets is shown as blue contours (image from Ylva
Pihlstrom, 1.29$'' \times 1.17''$ beam size). The radio
jet emission is anti-correlated with the X-ray emission of the corona. X-ray
emission from three radio knots E1, E2 and W1 (Fig. 1 of O'Dea \& Owen 1986),
is also detected. There are indentations in the east and west of the corona
along the jet direction, aligned with the knots. This indicates an
interaction between the jets and the X-ray gas.
b) The 0.5 - 1.5 keV \chandra\ contours superposed on the \hst\ WFPC2 image
(F702W filter). A dust lane, nearly perpendicular to the jet direction
and aligned with the major axis of the galaxy, is detected within the
central 1$''$ radius. A nearby bright star ($\sim 3''$ east of the nucleus)
distorts the optical light of the galaxy. The outer contours of the optical
light are also plotted. The X-ray light is asymmetric compared to the optical light.
c) \rosat\ contours (red) of the Perseus cluster superposed on the DSS I
image. The cluster center (NGC~1275) is on the lower left. A low resolution VLA image
of 3C~83.1B (blue contours) is shown (image from Chris O'Dea, Fig. 11 of
O'Dea \& Owen 1986), with NGC~1265 at the head of the NAT source. The tail of
the radio galaxy is at least 40$'$ in projection and bends sharply at least 4
times (see more figures in O'Dea \& Owen 1986 and Sijbring \& de Bruyn 1998).
The region around the galaxy (2.3$'\times2.3'$) is also enlarged on the left
insert, where contours of a 6cm VLA image (from Chris O'Dea) are superposed on
the DSS II image. The optical light of NGC~1265 can be traced to the edge of this
subfield.
   \label{fig:img:smo}}
\end{figure}

\begin{figure}
\vspace{-0.2cm}
  \centerline{\includegraphics[height=1.0\linewidth,angle=270]{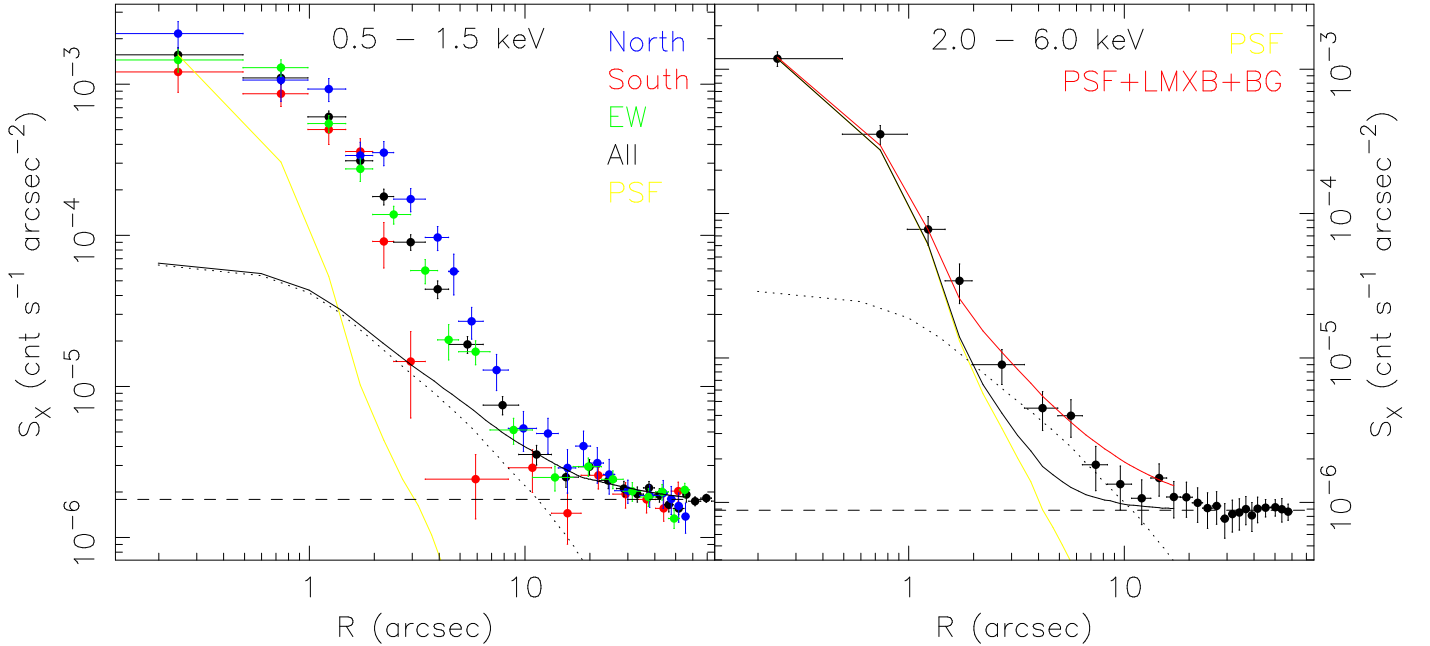}}
  \caption{{\bf Left}: the 0.5 - 1.5 keV surface brightness profiles
centered on the NGC~1265 nucleus (black: azimuthally averaged; red: south;
blue: north; green: east-west). The north region is from azimuthal angle
42\mydegree\ to 135\mydegree\ (measured from the west), while the south region
is from azimuthal angle 227\mydegree\ to 309\mydegree. The east-west region
includes all azimuthal angles left. The yellow line represents the local 0.5 -
1.5 keV PSF. The dashed line represents the 
background determined from the azimuthally averaged profile between 30$''$
and 120$''$. The dotted line is the optical light profile determined from the
\hst\ observations and is normalized to match the predicted LMXB luminosity
from the scaling relation by Sarazin et al. (2001),
with the assumption that the LMXB distribution follows the optical
light. The solid line is the sum of the previous two components.
The southern surface brightness profile shows a sharp edge at $\sim 3''$ (smoothed
by the PSF) from the nucleus, while the extension in the north can be traced
to at least 8$''$ (excluding the possible LMXB component).
{\bf Right}: the 2 - 6 keV surface brightness profile. The yellow line, dashed
line and the dotted line represent the same components as those in the left panel,
but in the 2 - 6 keV band. The black solid line is the sum of the PSF and
the local background, which clearly underestimates the data from 2$''$ to
8$''$. The red line is the sum of the PSF, the local background and the
LMXB component scaled from the optical light, which roughly reproduces
the data.
   \label{fig:img:smo}}
\end{figure}

\begin{figure}
  \includegraphics[height=0.8\linewidth]{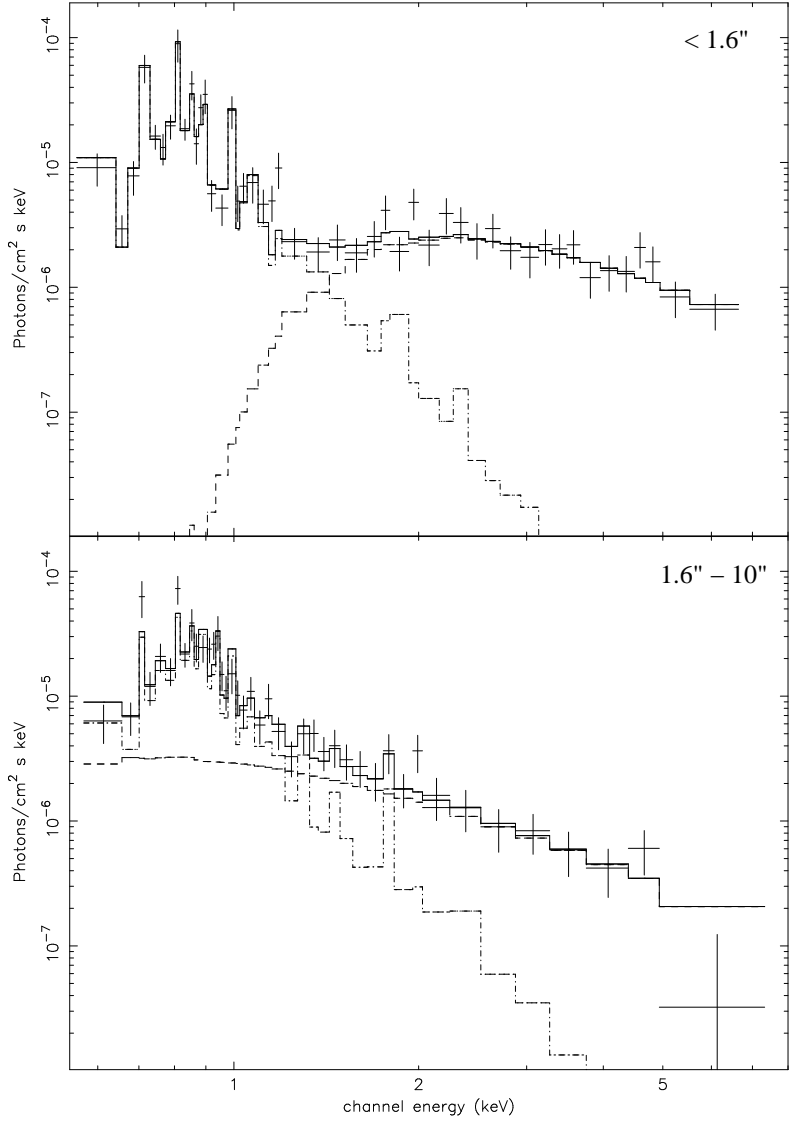}
  \caption{The spectra of the inner ($< 1.6''$) and outer (1.6$'' - 10''$)
regions of the NGC~1265 X-ray source, with the two-component models shown as
the fourth and the seventh rows in Table 1 (dot-dash line: corona emission;
dashed line: absorbed nuclear emission or LMXB emission; solid line: the sum).
The blend of iron L lines is significant in both spectra. The shapes of
the spectra at E $>$ 1.5 keV are different. The absorbed nucleus dominates
in the hard X-rays for the inner region, while the LMXB is a significant
component for the outer region.
   \label{fig:img:smo}}
\end{figure}

\begin{figure}
\vspace{-0.2cm}
  \centerline{\includegraphics[height=0.5\linewidth,angle=270]{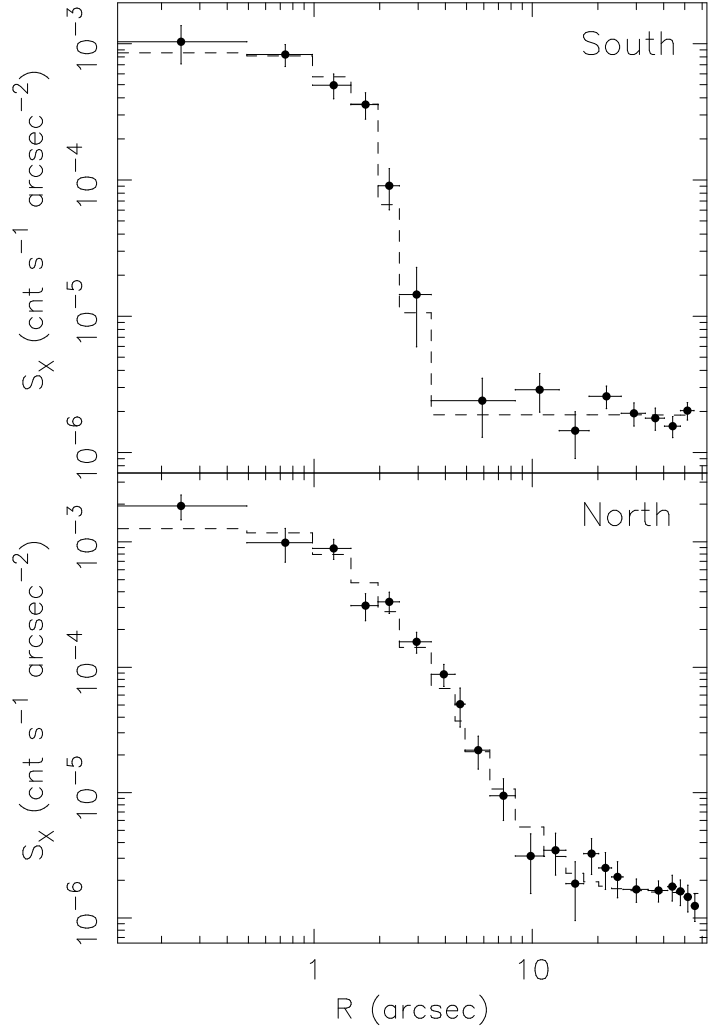}}
  \caption{{\bf Upper}: the 0.5 - 1.5 keV surface brightness profile
of the south sector with the best-fit model, a truncated $\beta$ model
with a constant background. The estimated nuclear emission at the 0.5
- 1.5 keV band is subtracted from the profile. The best-fit model has been
convolved with the local PSF, which smoothes the sharp edge. The surface
brightness jump across the edge is $\sim$ 106.
{\bf Lower}: the 0.5 - 1.5 keV surface brightness profile of the north
sector with the best-fit model, a $\beta$ model with a constant background.
The estimated nuclear emission and the LMXB emission is subtracted from the
profile. The best-fit model has been convolved with the local PSF.
The truncation of the $\beta$ model is not significant.
Despite the not so small errors, the gas distributions of these two sectors are
consistent with the same one, but with different truncation radii (Table 2).
   \label{fig:img:smo}}
\end{figure}

\end{document}